# Pulsed laser deposition of highly c-axis oriented thin films of BSTS topological insulator

Atul Pandey[1], Sourabh Singh[2], Bishnupada Ghosh[1], Subhadip Manna[1], R. K. Gopal[3,*], and Chiranjib Mitra[1,*]

[1] Department of Physical Sciences, Indian Institute of Science Education and Research Kolkata, India
[2] Racah Institute of Physics, The Hebrew University of Jerusalem, Israel
[3] Department of Physical Sciences, Indian Institute of Science Education and Research Mohali, India
*Corresponding Authors: chiranjib@iiserkol.ac.in, gopalrk7@gmail.com

We report the growth of highly c-axis oriented topological insulator (TI) $BiSbTe_{1.5}Se_{1.5}$ (BSTS) thin films by pulsed laser deposition (PLD) technique. The various growth parameters such as substrate temperature, Argon pressure in the deposition chamber and target to substrate distance are tuned to obtain the optimized conditions essential for stoichiometric and bulk insulating TI thin films. These films are highly c-axis oriented and exhibit all the four Raman modes characteristic to the $R\bar{3}m$ space group. The quality of the deposited thin films is investigated using X-ray diffraction for crystallinity, Raman spectroscopy for lattice dynamics, morphological studies using scanning electron microscope and compositional analysis using Energy-dispersive X-ray spectroscopy.  Resistance vs temperature measurements confirm bulk insulating nature of the prepared thin films and magnetoresistance data exhibits the phenomena of weak antilocalization with a large phase coherence length.

## I. INTRODUCTION

Three-dimensional Topological Insulators (TIs) are a unique class of quantum materials having an insulating bulk and linearly dispersed spin-polarized gapless surface states [1]. The topological protection of the metallic surface states owing to the spin-momentum locking provides them protection from backscattering against non-magnetic defects and impurities. [2-3]. These interesting attributes make TIs not just an interesting state of matter but also immensely promising in a plethora of fields ranging from spintronics [4-5], quantum computing [6-7], etc. Such novel applications of the TIs require the dominance of the surface transport in comparison to bulk transport, achieved by having highly insulting bulk.

Recently, many works reported the thin film growth of binary TI materials by pulsed laser deposition (PLD) [8-13]. However, a highly insulating bulk has rarely been achieved in prototypical binary TIs, because of naturally occurring defects and the resulting unwanted doping [14-16].The second-generation TIs such as $Bi_2Se_3$ (BS), $Bi_2Te_3$ (BT) and $Sb_2Te_3$ (ST) all suffer from parasitic bulk defects, which drives the chemical potential to either the conduction or valence band [17-18]. An essential requirement to obtain the intrinsic TI regime is to quench these defect states and thus reducing the contribution of the trivial bulk states in the transport measurements [19]. In this regard, solid solution tetradymite compound $Bi_{2-x}Sb_xTe_{3-y}Se_y$ (BSTS), for a special combination of x and y values were shown by transport measurements to be one of the most promising candidates for highly insulating bulk conduction [20-21]. The idea behind this was that the two opposing types of charged defect: (Bi, Sb)/ Te antisite defects and the Se vacancy defects compensate each other and effectively reduce the total bulk conductivity. Therefore, it is imperative to optimize the stoichiometry of the BSTS compound which in turn controls the positioning of the chemical potential in the bulk bandgap.

Recently, Ham *et al*. reported the growth of single crystals of BSTS by three different methods [22] and outlined that the stoichiometry of the compound is crucial to control the chemical potential of the surface states to be within the bulk gap. Despite recent studies, the single crystal growth of stoichiometric BSTS, containing multiple volatile elements like Se and Te remains a challenging task [21-24]. Apart from reducing the bulk conduction by proper chemical doping and stoichiometry control [21, 25-26] another effective way to achieve surface dominated transport is to increase the surface to bulk ratio by synthesizing thin films or nanostructure [27]. Thin films of the TIs have fewer numbers of point defects relative to the bulk of these materials [28-29]. Thus, the thin films have the potential to have idealistic surface dominant transport. Also, for thermoelectric applications of TI materials, thin films have distinct advantages [8]. Keeping this in mind we utilize the excellent property of PLD technique of stoichiometric transfer to prepare thin films of BiSbTe$_{1.5}$Se$_{1.5}$ using an alloy target of the same composition. Even though Se and Te are volatile materials, in this report, we explicitly show that an optimized set of deposition conditions can lead to an exact stoichiometric transfer from the target to the sample. We show that at high deposition pressure (~ 0.6 mbar) and appropriate temperature window, stoichiometric deposition of BSTS thin film is possible.

While Ngoc *et al*. reported the growth of thin nanoplates BSTS using a catalyst-free vapor solid method [30]. We report the successful growth of highly c-axis oriented and stoichiometric thin films of topological insulator BSTS by PLD. To the best of our knowledge, this is the first work on a systematic study of the growth of bulk insulating quaternary tetradymite stoichiometric BSTS thin films by PLD technique. Also, most of the electronic devices are multi-layered heterostructure; therefore, for the device application of TIs, a flexible thin film deposition technique is required. Multicomponent layered heterostructure can be easily prepared by PLD because of the ease of changing the source materials and combining multiple targets [31].

## II.  EXPERIMENTAL

BSTS thin films were deposited on Silicon Si (111) substrates by PLD in a flowing argon environment. To prepare the target 1:1 molar ratio of Bi$_2$Te$_3$ (Alfa aesar 99.98% purity) and Sb$_2$Se$_3$ (Alfa aesar 99.98% purity) were continuously grinded using a mortar and pestle in the acetone solution for 10 hours. The grinding time is a crucial factor in obtaining a uniform and good target. We earlier tried with another target that was prepared by grinding for only four hour and we did not obtain desired results. Prior to deposition the substrates were thoroughly and sequentially cleaned with HPLC grade acetone, methanol and DI water followed by drying with nitrogen and then immediately loaded into the deposition chamber. The chamber was evacuated to a base pressure of $5 \times 10^{-6}$ mbar before ablating the target with 1500 pulses of KrF excimer laser (wavelength: 248 nm and pulse width of 25 ns) at a repletion rate of 1 Hz. The spot size of the laser on the target was measured to be roughly 2 mm$^2$, and the energy of the laser pulse was 50 mJ; therefore the energy density being roughly 2.5 J/cm$^2$. To study the effect of substrate temperature (T$_S$), we deposited thin films at a fixed background Ar pressure (P$_{Ar}$ = 0.6 mbar). The films were prepared by two step growth process, where first 250 pulses were deposited at substrate temperature 140° C and the remaining 1350 pulses at T$_S$ = 120°C, 150°C, 180°C, 220°C, 230°C, 240°C, 250°C, 260°C, 280°C, 290°C, 300°C, 320°C and 380°C. The two-step technique has already been shown to be a better technique for growing thin films [32]. We have also compared the quality of the thin film deposited by two-step process with the film deposited by one step process, where at optimized T$_S$=250 and P$_{Ar}$= 0.6, all the 1500 pulses were fired at the target. To find the optimized Ar pressure we fixed $T_S$= 250°C and varied the $P_{Ar}$ = 0.18 mbar, 0.33 mbar, 0.45 mbar, 0.53 mbar, 0.6 mbar, and 0.68 mbar. Morphological studies of the grown film were performed by a Scanning Electron Microscope (SEM) (Carl Zeiss, Supra 55VP), for the compositional study a 15×15 µm$^2$ area was scanned using energy dispersive spectroscopy (EDS) attached to SEM with the acceleration voltage set to 10 kV. The vibrational dynamics study of the as-grown film was carried out using a mono vista CRS+ micro Raman spectrometer. The excitation wavelength used for the Raman measurements was 633 nm, and backscattering measurement configuration was used. The recorded spectra were collected using an objective lens of 20X magnification and the 1500 lines/mm grating. A spot size of roughly 10 microns was used to excite the sample. Crystallinity characterizations were done using XRD measurement in a Smart lab Rigaku machine

## III. RESULTS
### A. Morphology and composition

SEM images of some selected thin films grown at different deposition conditions are shown in figure (1). Thin films deposited using the two-step process, for the temperature window of 120°C < $T_S$ < 240°C at fixed $P_{Ar}$ =0.6 mbar, and for 0.33 mbar <$P_{Ar}$< 0.68 mbar at fixed $T_S$=250°C are granular in nature, featuring hexagonal like crystallites. Granularity increases with decrease of $T_S$ and increase of $P_{Ar}$. Average grain size as estimated from the SEM images using ImageJ software shows an increasing trend with increase of $T_S$. At $T_S \geq 280$°C, grain size increases beyond the distance between and neighboring grains and different grains overlap. The dependence of grain size and morphology on $T_S$ and $P_{Ar}$ comes through the dependence of mobility of the adatoms on $T_S$ and $P_{Ar}$. The Ar absorbed between the grain boundaries tends to suppress grains at higher $P_{Ar}$ mobility increases with increase of $T_S$ [8]. Films grown by the standard one step process at optimized $T_S$ (250°C) and $P_{Ar}$ (0.6 mbar) does not exhibit granular nature as shown in figure 1 (J). This is an important observation as granularity is an important attribute of TI thin films. It has been shown that granular thin films exhibit interesting attributes and have huge

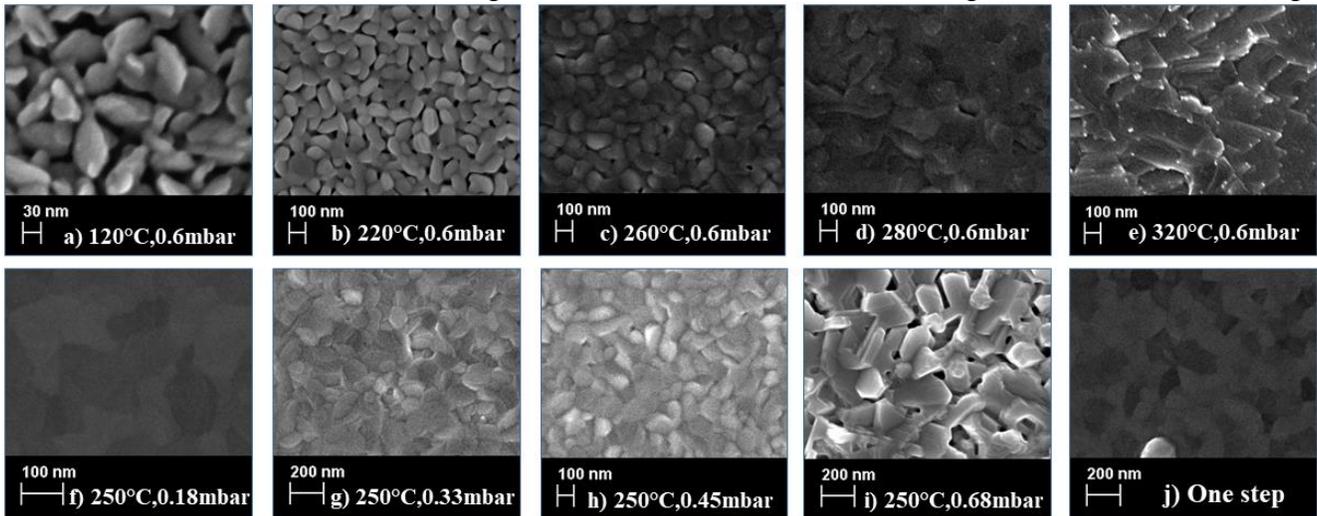

practical implications [33].

*Figure.1: SEM Images of films deposited at different mentioned parameters. Figure a)-i) shows films deposited following the two-step process, figure j) shows the film deposited following the one step process*

One of the primary concerns in preparing these quaternary thin films is to control the volatile elements Se and Te and maintain the requisite stoichiometry. Thus, it is necessary to measure the stoichiometry of our prepared thin films. The Bi:Sb:Te:Se ratio of the selected films as determined from EDS measurements and are shown in Table 1. All the results at $T_S=240°C$, $P_{Ar}=0.6$ mbar is within an error of 10%, which is comparable with previous reports [10, 22]. The film deposited at $T_S = 320°C$ shows reduced Te and Sb content ($Bi_{1.47}Sb_{0.42}Te_{1.05}Se_{2.04}$), the reason for which is still unknown. The target to substrate distance is another key parameter for obtaining stoichiometric TI thin films [10], the optimized distance for our thin films was obtained to be 4.7 cm.

| $P_{Ar}=0.6$ mbar | | | $T_S=250°C$ | |
|---|---|---|---|---|
| $T_S=180°C$ | $T_S=240°C$ | $320°C$ | $P_{Ar}=0.18$ mbar | $P_{Ar}=0.45$ mbar |
| 1.04:1.06:1.34:1.54 | 1.1:0.9:1.46:1.53 | 1.47:0.42:1.06:2.05 | 0.91:0.85:1.37:1.87 | 1:0.9:1.37:1.7 |

*Table 1. Energy Dispersive X-ray Spectroscopy (EDS) analysis of BSTS thin films deposited at different growth parameters*

## B. Crystallinity

The crystal structure of the BSTS is a rhombohedral belonging to the space group ($R\bar{3}m$), which can be thought as periodic repetition of -A-B-C- layers. Thus, the crystal exhibits a layered structure that can be grouped into a set of five monolayers referred to as the quintuple layer (QL) as shown in the figure 2(b) and (c). The five monolayers are of the form $A^{VI}(1)$ - $B^V$ - $A^{VI}(2)$ - $B^V$ - $A^{VI}(1)$, where $A^{VI}$ and $B^V$ are elements from the periodic table group VI and V respectively and the number in the parenthesis of $A^{VI}$ is to differentiate two different lattice sites occupied by $A^{VI}$. In the case of BSTS, $A^{VI}$ can be Se or Te and $B^V$ can be Bi or Sb. Because of large electronegativity of Se compared to Te; Se occupies $A^{VI}(2)$ site, remaining 0.5 Se and 1.5 Te occupies $A^{VI}(1)$ sites [34].

XRD patterns after subtracting the substrates peaks are shown in the figure (3). For the films grown at $T_S = 240°C$ to $280°C$, XRD pattern matches exactly with the previous reports of the BSTS single crystals of similar composition [22] and with C-axis oriented polycrystalline thin films of other TIs [10-12,35-40]. Thus, only (0 0 3n) peaks are present in the films except the (0 0 9) peak, which got suppressed because of strong substrate peak, indicating the preferential c-axis orientation along the [111] silicon substrate crystallographic direction. For the films grown below 240°C few (0 0 3n) peaks starts to disappear and additional peak other than (0 0 3n) appears and peak intensity to background noise ratio decreases significantly. This implies that $T_S<240°C$ is insufficient for atomic migration and establishing an epitaxial film-substrate interface. Films grown at 320°C exhibits broad (0 0 6) and (0 0 15) peaks with shoulder, such diffraction pattern was also observed previously by Ham *et al*. [22] for the single crystal. They claimed it was due to the inhomogeneous composition resulting in anti-site defects or vacancies. Film deposited at 380°C shows only substrates peaks, indicating amorphous nature. As shown in figure 3, films deposited at $P_{Ar}$ in the range of 0.18 mbar to 0.68 mbar exhibit c-axis preferred orientation.

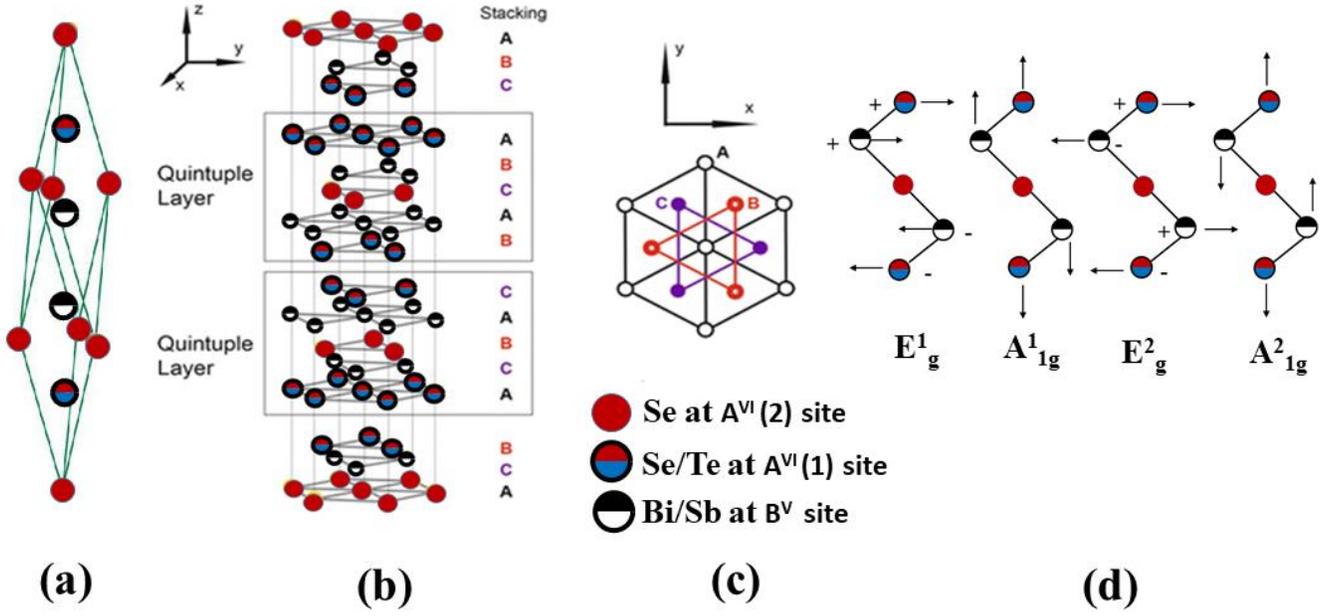

*Crystal structure of BSTS a) In the primitive rhomboidal unit cell. b) In the hexagonal unit cell. c) Top tiew of layered structure. d) Raman active vibrational modes.*

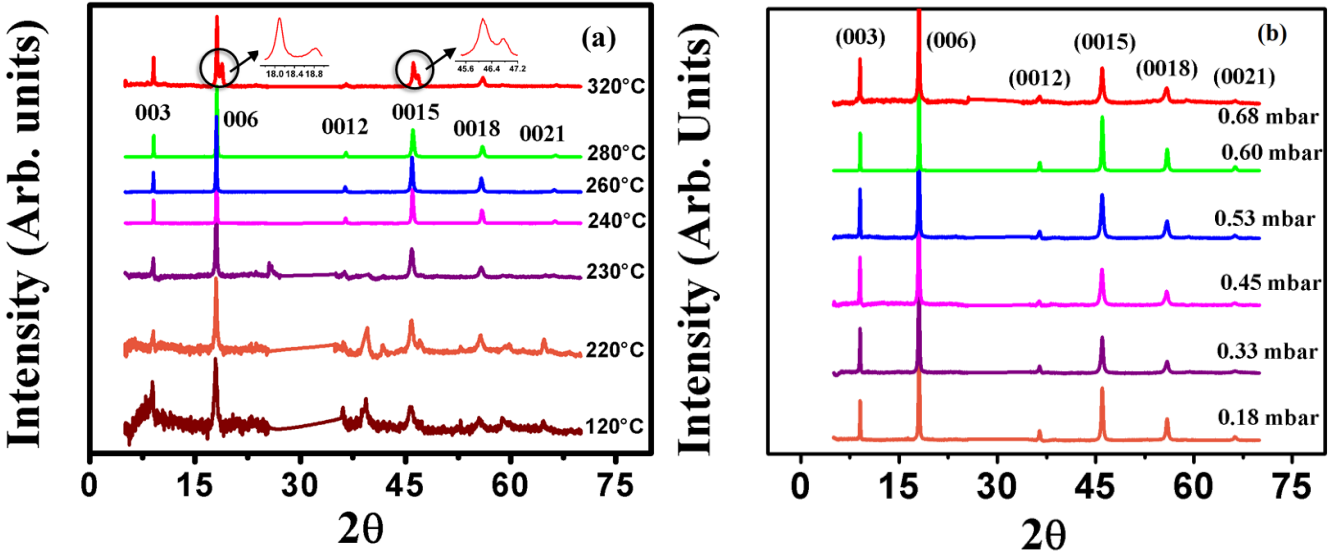

*Figure 3. (a) X-ray diffraction pattern of BSTS thin film deposited at different substrate temperature keeping the Ar background pressure fixed at 0.6 mbar. (b) X-ray diffraction pattern for the BSTS thin films deposited at varying Ar background pressure.*

Using the following formula

$$\lambda = 1.5406 \text{ Å} = 2d\sin(\theta) = 2\frac{c}{l}\sin(\theta_l) \tag{1}$$

For (0 0 l) peak we calculated the lattice constant *c*. As shown in the figure 4 (a). It decreases from 29.7Å for $T_s$ = 120°C to 29.38Ä for $T_s$ = 320°C and remains roughly constant ≈ 29.45Å for different $P_{Ar}$. Value of c for the TI of similar composition, $BiSbTeSe_2$ single crystal is 29.64 Å [22]

We analyze Full width at Half Maxima (FWHM) of the highest intensity peak (0 0 6), as shown in figure 4(b). FWHM decreases with increase of $T_S$ and remains roughly constant with respect to $P_{Ar}$. For 240°C≤$T_S$≤320°C and 0.18 mbar ≤ $P_{Ar}$ ≤ 0.68 mbar, FWHM remains less than 0.2 which is close to that for BS thin films grown by PLD [10-12] and other TI thin films prepared by MBE [41-42].

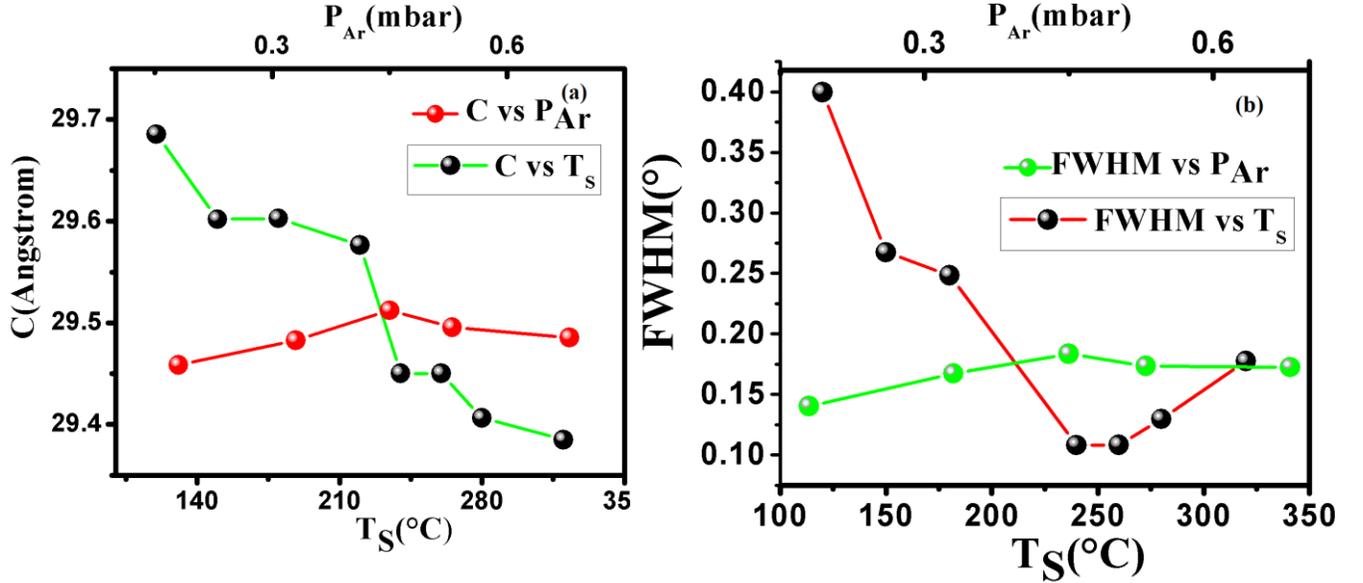

*Figure 4(a) Lattice constant "c" as a function of $P_{Ar}$ and $T_S$. b) FWHM as the function of $P_{Ar}$ and $T_S$.*

## C. Dynamical properties

There are five atoms in the primitive unit cell of the BSTS; hence, there are 15 dynamical vibrational modes. Of these three are acoustic phonon and 12 are optical phonon modes. 6 out of the 12 phonon modes are Raman active. At the Brillouin zone center, the irreducible representations for six Raman, active modes can be written as $2A_{1g} + 2E_g$. Two out of plane vibration as shown in the figure (2.d) are designated as $A_{1g}^1$ (low frequency out of xy plane vibration) and $A_{1g}^2$ (high frequency out of xy plane vibration). There are two different degenerate in plane vibrational modes designated as $E_g^1$ (low frequency in plane vibration) and $E_g^2$ (high frequency in plane vibration. Richter *et al.* investigated the phonon modes in the rhombohedral $A_2B_3$ crystals [43]. They used a simple model to argue that $B^V$ - $A^{VI}(2)$ bonding forces are primarily involved in the $E_g^1$ and $A_{1g}^1$ modes. For the $E_g^2$ and $A_{1g}^2$ modes, $A^{VI}(1)$ and $B^V$ move in opposite directions, the bonding forces $A^{VI}(1)$ - $B^V$ would mainly affect the $E_g^2$ and $A_{1g}^2$ modes frequencies. If the bonding forces $B^V$ - $A^{VI}(2)$ are smaller than the $A^{VI}(1)$ - $B^V$ bonding forces [43], normal mode frequencies are in the order $E_g^2 > E_g^1$ and $A_{1g}^2 > A_{1g}^1$. Jian Yuan *et al.* [44] argued that $A_{1g}$ modes which undergoes out of plane vibration, induces shorter displacement of vibrating atoms compared to the in-plane vibrations of $E_g$ modes. A short displacement induces higher phonon frequencies; consequently, $A_{1g}$ modes vibrate at higher frequency than the corresponding $E_g$ modes. Hence the four Raman normal mode frequencies are in the order of $A_{1g}^2 > E_g^2 > A_{1g}^1 > E_g^1$.

Richter *et al.* observed three of the four normal mode frequencies for pure $Bi_2Se_3$ crystal at the following wavenumber 72 cm$^{-1}$ ($A_{1g}^1$), 131.5 cm$^{-1}$ ($E_g^2$) and 174.5 cm$^{-1}$ ($A_{1g}^2$) [43]. Jun Zhang *et al.* [45] for the first time observed all the four Raman peaks in $Bi_2Se_3$ nanoplates at 37 cm$^{-1}$ ($E_g^1$), 72 cm$^{-1}$ ($A_{1g}^1$), 131.5 cm$^{-1}$ ($E_g^2$) and 174.5 cm$^{-1}$ ($A_{1g}^2$). Other Raman studies on the TIs of the composition $Bi_xSb_{2-x}Te_3$, $Bi_xSb_{2-x}Se_3$, $Bi_2Te_ySe_{3-y}$, and $Sb_2Te_ySe_{3-y}$ showed that frequency of the $E_g^2$ and $A_{1g}^2$ modes decreases significantly with increase of Te content and decrease of Sb content [37,44,46-47].

Raman spectra of films deposited at different parameters are shown in the figure (5). We observe all four expected Raman modes at frequencies ~38.5 cm$^{-1}$ ($E_g^1$), ~ 68 cm$^{-1}$ ($A_{1g}^1$), ~ 113 cm$^{-1}$ ($E_g^2$) and ~ 162 cm$^{-1}$ ($A_{1g}^2$). We compare our result with recent work of German *et al.* who calculated normal mode frequencies using Density Functional Theory for a BSTS TI of very similar composition $BiSbTeSe_2$ [34]. They obtained four Raman normal mode frequencies at 38.5 ($E_g^1$), 69 ($A_{1g}^1$), 114.5 ($E_g^2$) and 180.1 ($A_{1g}^2$), which is very close to our results, except $E_g^2$ and $A_{1g}^2$ modes. This anomaly arises because of the compositional difference between their sample $BiSbTeSe_2$ and our compound $BiSbTe_{1.5}Se_{1.5}$. Therefore, due to the higher Selenium content in our sample, frequency of the $E_g^2$ and $A_{1g}^2$ modes is slightly higher. Decreasing the $P_{Ar}$, or increasing $T_S$ results in reaching our optimized deposition condition the morphology becomes compact and subsequently leads to broadening of the $A_{1g}^2$ mode. This is an interesting observation and similar behavior was observed in nanoplates, which exhibited broadening of this mode with increasing thickness [48, 49].

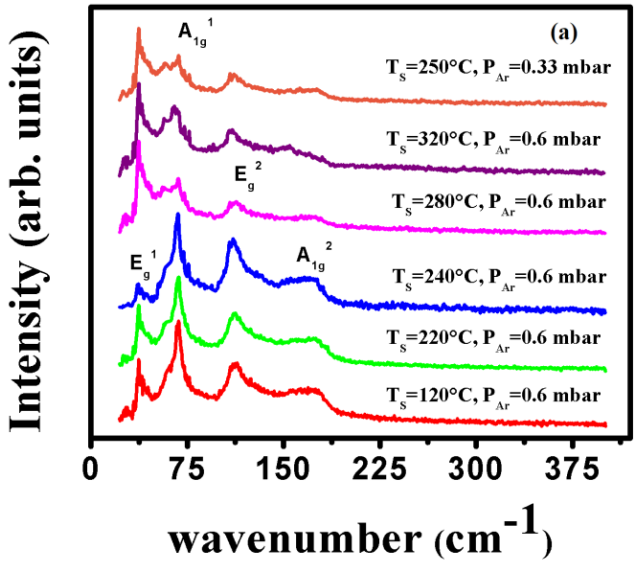
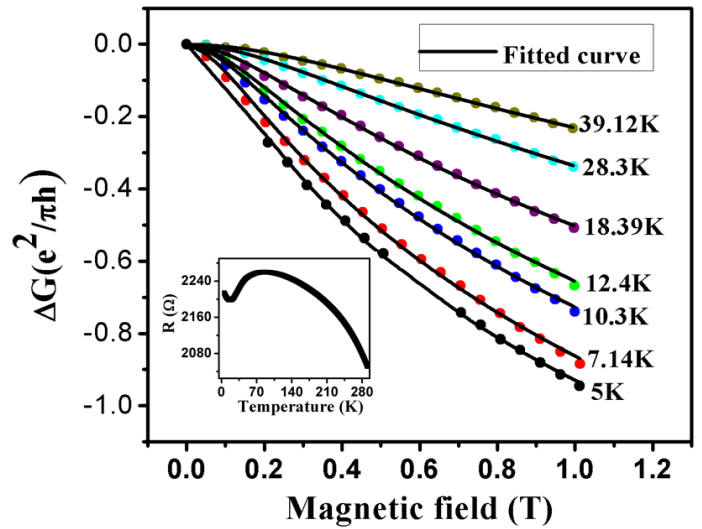

*Figure 5. Raman spectra of BSTS thin films deposited at different parameters.*

*Figure 6. HLN fitting of the Magneto conductance data of the BSTS thin film. The Resistance vs. temperature data of the film is shown in the inset.*

## D. Transport properties

Transport measurements are performed for the film grown at optimized parameter of $T_S=260°C$ and $P_{Ar}=0.6$ mbar. The measurements were performed using a cryogen free magnet from Cryogenic U.K. system from room temperature down to 5 K. Temperature dependence of the resistance as shown in the inset of the figure (6) shows bulk insulating behavior in agreement to previous reports on BSTS thin films and single crystals [21,32]. The magneto conductance ($\Delta G$) data is fitted to simplified Hikami, Larkin and Nagaoka (HLN) [50]

$$\Delta G = \frac{w}{l}\sigma = -\frac{w}{l}\alpha\frac{e^2}{\pi h}\left[\psi\left(\frac{\hbar}{4e^2 L_\phi^2 B}+\frac{1}{2}\right) - \ln\left(\frac{4e^2 L_\phi^2}{\hbar B}\right)\right] \qquad (2)$$

Where, $w$ and $l$ respectively are the length and width of the sample, $\alpha$ is a constant related number of conducting channels, and $L_\phi$ is the phase coherence length. The magnetoconductance data and the fitted curve are shown in figure (6). The extracted parameter $\alpha$ remains close to 1 in the temperature range of 5-40 K which is an indication of two independent conducting channels [51]

## IV. DISCUSSION

We observe that the films deposited in the temperature window of 240°C $T_S \leq$ 280°C exhibit c-axis preferentially oriented. Substrate temperature below 240°C is insufficient to establish preferential growth, and at higher temperature films tend to be amorphous. In previous reports of TI thin films deposited by PLD, obtained optimized temperature range for epitaxial growth are slightly different: 300°C - 350°C for BT [11], 280°C-300°C [10], 300°C-350°C[9] for BS and 200°C-400°C for ST [12]. However, many other parameters such as substrate used, laser energy density, the argon background pressure etc. were also different in those studies.

In earlier works of characterizing PLD grown thin film of binary and ternary TIs and BSTS thin films growth by other deposition methods [9-14, 30], most of the studies focused either on vibrational or crystallinity studies. With combined vibrational and crystallinity studies we see that growth temperature affects the phase of the films, at low deposition temperature, though the films are not preferentially oriented, there is still formation of BSTS granular films as revealed by Raman data and morphology of the films. It would be interesting to do transport studies on such textured BSTS thin films and see how phase coherence length is affected.

In a wide Argon pressure range of 0.18 mbar to 0.68 mbar, films grown are highly c-axis oriented. The argon pressure mainly changed the granularity of the deposited thin films with granularity increasing with increase of argon pressure. In our study we find that at 0.68 mbar of background Ar pressure, the plume length decreases significantly and consequently the particulates cannot reach the substrate kept at optimized distance of 4.7 cm and hence the pressure cannot be increased further. Similarly, at lower pressure of 0.18m mbar plume is too much spread, which leads to uneven deposition in terms of stoichiometry. In the pressure range of 0.45 mbar to 0.6 mbar, plume shows ideal ellipse like shape, where all particulates terminate at point and the substrate can be placed at this terminating point.

## V. SUMMARY

We deposited TI thin films using PLD technique and optimized the various deposition conditions to define the favorable substrate temperature window and pressure conditions to obtain stoichiometric and high quality BSTS thin films. XRD measurements confirm c-axis oriented thin films and the performed Raman spectroscopy exhibit all four peaks pertaining to the various modes. Transport data confirms bulk insulating behavior and the weak antilocalization data is fitted to HLN equation. . To summarize, high quality and stoichiometric bulk insulating TI thin films can be prepared using PLD method and the variation of the deposition conditions renders us a tool to prepare granular high-quality thin films.

## ACKNOWLEDGEMENTS

Authors would like to acknowledge Ministry of Human Resource and Development (MHRD), Govt. of India for the funding, and IISER Kolkata for providing experimental facilities.